\documentclass[a4paper]{article}
\usepackage{Odyssey2022}
\usepackage{epsfig,amssymb,amsmath}
\usepackage{url,cite}
\usepackage{hyperref}
\usepackage{xpatch,xcolor}
\usepackage{multicol, multirow}
\usepackage{pifont}
\usepackage{tabularx}

\hypersetup{
    colorlinks=true,
    linkcolor=purple,
    filecolor=magenta,      
    urlcolor=purple,
}

\newcolumntype{Y}{>{\centering\arraybackslash}X}

\usepackage{footnote}
\usepackage[symbol]{footmisc}

\newcommand{\newpara}[1]{\vspace{6pt}\noindent\textbf{#1}}

\title{Baseline Systems for the First Spoofing-Aware Speaker Verification Challenge:\\Score and Embedding Fusion}
\name{\parbox{\linewidth}{\centering
Hye-jin Shim$^{*,1}$\thanks{$^{*}$These authors contributed equally to this work.}, Hemlata Tak$^{*,2}$, Xuechen Liu$^{3,8}$, Hee-Soo Heo$^{4}$, Jee-weon Jung$^{4}$
Joon Son Chung$^{5}$,\\ Soo-Whan Chung$^{4}$, Ha-Jin Yu$^{1}$, Bong-Jin Lee$^{4}$, Massimiliano Todisco$^{2}$ Héctor Delgado$^{6}$,\\ Kong Aik Lee$^{7}$, Md Sahidullah$^{8}$, Tomi Kinnunen$^{3}$, Nicholas Evans$^{2}$}}
\address{
  $^1$School of Computer Science, University of Seoul, South Korea\\
  $^2$EURECOM, Sophia Antipolis, France, $^3$University of Eastern Finland, Finland\\
  $^4$Naver Corporation, South Korea, $^5$KAIST, South Korea\\
  $^6$Nuance Communications, Spain, $^7$A$^*$STAR, Singapore, $^8$Inria, France\\
  {\small \tt sasv.challenge@gmail.com}
  }

\begin{document}
\maketitle
\begin{abstract}
Deep learning has brought impressive progress in the study of both automatic speaker verification (ASV) and spoofing countermeasures (CM).
Although solutions are mutually dependent, they have typically evolved as standalone sub-systems whereby CM solutions are usually designed for a fixed ASV system.
The work reported in this paper aims to gauge the improvements in reliability that can be gained from their closer integration. 
Results derived using the popular ASVspoof2019 dataset indicate that the equal error rate (EER) of a state-of-the-art ASV system degrades from 1.63\% to 23.83\% when the evaluation protocol is extended with spoofed trials.
However, even the straightforward integration of ASV and CM systems in the form of score-sum and deep neural network-based fusion strategies reduce the EER to 1.71\% and 6.37\%, respectively.
The new Spoofing-Aware Speaker Verification (SASV) challenge has been formed to encourage greater attention to the integration of ASV and CM systems as well as to provide a means to benchmark different solutions.
\end{abstract}
\textbf{Keywords}: automatic speaker verification, anti-spoofing, spoofing-aware speaker verification, spoofing countermeasures.

\section{Introduction}
\label{sec:intro}

Recent years have seen rapid progress in automatic speaker verification (ASV)~\cite{snyder2018x, desplanques2020ecapa,bai2021speaker}. 
Even for unconstrained \textit{in the wild} scenarios, the latest systems deliver low equal error rates (EERs) that are close to those for well-constrained conditions~\cite{garcia2020magneto,desplanques2020ecapa,thienpondt2021integrating}.
However, there is evidence that these improvements might not offer protection against \emph{spoofing attacks} -- the presentation of utterances specially crafted to deceive the ASV system.

Solutions to protect ASV systems from such attacks take the form of countermeasures (CMs), typically separate sub-systems designed to detect manipulated or synthetic utterances~\cite{nautsch2021asvspoof}.
The threat of spoofing attacks has intensified in recent times due to the rapid advances in other speech technologies which can be used to generate spoofed utterances.  
They include: speech-to-speech voice conversion (VC); text-to-speech (TTS) speech synthesis; replay attacks.
Since ASV systems are increasingly deployed in security-critical operations as a part of a biometric authentication system, vulnerabilities to spoofing attacks are unacceptable.

In response to the threat, the ASVspoof initiative has held biennial challenges to promote the development of research in spoofing detection~\cite{nautsch2021asvspoof}. 
Two different use case scenarios have been defined, namely physical access (PA) and logical access (LA).
The work in this paper relates to the latter, 
typically telephony applications and robustness to TTS and VC spoofing attacks. When assessed using the ASVspoof 2019 LA evaluation set, today's leading CM systems deliver EERs of less than 2\%~\cite{lavrentyeva2019stc,chen2020generalization,tak2021end,hua2021towards,li2021channelwise,wang2021comparative,zhang2021effect,luo2021capsule,ge2021raw,jung2022aasist,wang2022practical}.

While the EER metric was adopted in almost all early work, the ASVspoof community has now transitioned to the minimum tandem detection cost function (min t-DCF)~\cite{kinnunen-tDCF-TASLP} as the primary metric.
The t-DCF reflects the impact of spoofing and countermeasures upon a typically-fixed ASV system.
Even with this strategy, CMs are often designed in standalone fashion, independently from ASV.
Until now, and with only few notable exceptions~\cite{sahidullah2016integrated,todisco2018integrated, li2019multi,li2020joint,sizov2015joint, shim2020integrated,gomez2020joint,kanervisto2021optimizing}, very little work has investigated the benefit of jointly optimised, or integrated CM+ASV solutions. 



The \textbf{Spoofing-Aware Automatic Speaker Verification} (SASV) challenge\footnote{\url{https://sasv-challenge.github.io}}, a special session at INTERSPEECH~2022, aims to promote greater research in this direction and extends the traditional ASV scenario to consider spoofing attacks.
The first SASV challenge~\cite{jung2022sasv} utilises existing ASVspoof databases with metrics extended to support the evaluation of integrated CM+ASV solutions.
Ultimately, SASV aims to strengthen the foundations between research in spoofing detection and ASV.

New contributions reported in this paper include:
(i)~baseline SASV solutions to integrated CM+ASV leveraging state-of-the-art sub-systems;
(ii)~metrics designed specifically for the SASV task; 
(iii)~experimental results and detailed, per-attack analyses. 


\section{Related work}
\label{sec:related_works}
\begin{figure}[!t]
  \begin{center}
    \centering
    \includegraphics[trim={5cm 6.5cm 3cm 0.0cm},clip,width=1.475\linewidth]{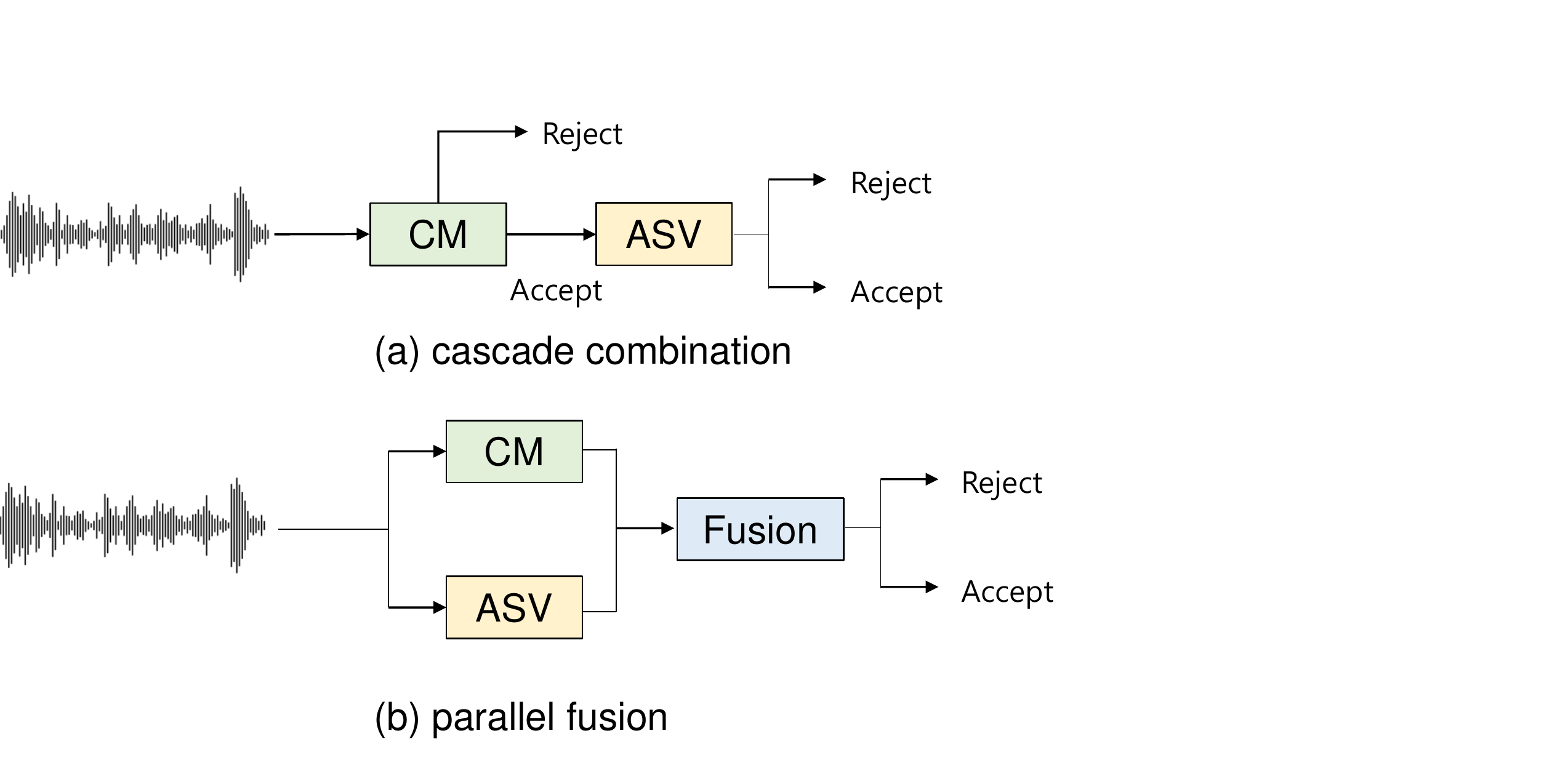}
    \caption{Back-end fusion of CM and ASV sub-systems. (a): cascaded combination, a form of decision level fusion. (b): parallel fusion which can operate at either decision, score, or embedding levels. When parallel fusion operates at the decision level, it is the same as the cascaded combination.}
    \label{fig:three_cascade}
  \end{center}
\end{figure}

The majority of previous related work focuses upon the development of independent CM and ASV solutions. 
Comparatively very little work has explored their integration.
The literature can be divided into two strands: (i) back-end fusion using independent ASV and CM sub-systems and (ii) single SASV systems.

The earlier works in the first strand investigated decision-level cascade or parallel combinations~\cite{khoury2014introducing, sahidullah2016integrated, todisco2018integrated}.
As shown in Fig.~\ref{fig:three_cascade}-(a), the cascaded combination typically involves the use of a CM as a gate prior to ASV so that the latter treats only input utterances labelled by the CM as bona fide.
It can be regarded as a form of decision level fusion.
For parallel combinations, as illustrated in Fig.~\ref{fig:three_cascade}-(b), every input utterance is treated by both sub-systems before the outputs are fused. 
Fusion can be performed at decision, score or embedding levels.
Given decision level fusion and identical CM and ASV thresholds, cascade and parallel solutions give identical results.

Beyond straightforward score fusion, \cite{todisco2018integrated} reports a Gaussian back-end fusion strategy with different front-ends for CM and ASV sub-systems. 
The Gaussian back-end fusion method is used to model ASV and CM scores as two-dimensional vectors from which single scores are derived.
The Gaussian back-end is shown to outperform both cascade and straightforward score fusion parallel combination strategies by a large margin. 

The second strand of single SASV systems has been also explored~\cite{li2019multi,li2020joint,shim2020integrated,gomez2020joint}.
An approach to the joint training of CM and ASV systems using multi-task learning (MTL)~\cite{caruana1997multitask} is reported in~\cite{li2019multi, li2020joint}. 
However, the framework requires DNN training towards the speakers in the enrolment set and cannot incorporate new speakers, making the framework somewhat inflexible.
The first single SASV system adaptable towards unlimited speakers is reported in~\cite{shim2020integrated}.
However, the single SASV system is outperformed by a back-end DNN fusion approach (the former strand) by a large margin. 

While some of the above referenced works were performed with standard databases, none used common protocols to assess ensemble or integrated CM+ASV approaches.
The absence of benchmarking frameworks means that
results for different approaches cannot be compared meaningfully and hinders the development of integrated systems.
The SASV challenge has been designed to address these issues, to establish such a common benchmarking framework and to promote progress in the integration and joint optimisation of CM+ASV solutions.
The remainder of this paper introduces our baseline systems, protocol, metrics, and results.  

\section{Embedding extraction}
\label{sec:embd_extract}

\begin{figure*}[!ht]
  \centering
 \includegraphics[trim={0cm 0cm 3.7cm 0.1cm},clip,width=0.85\linewidth]{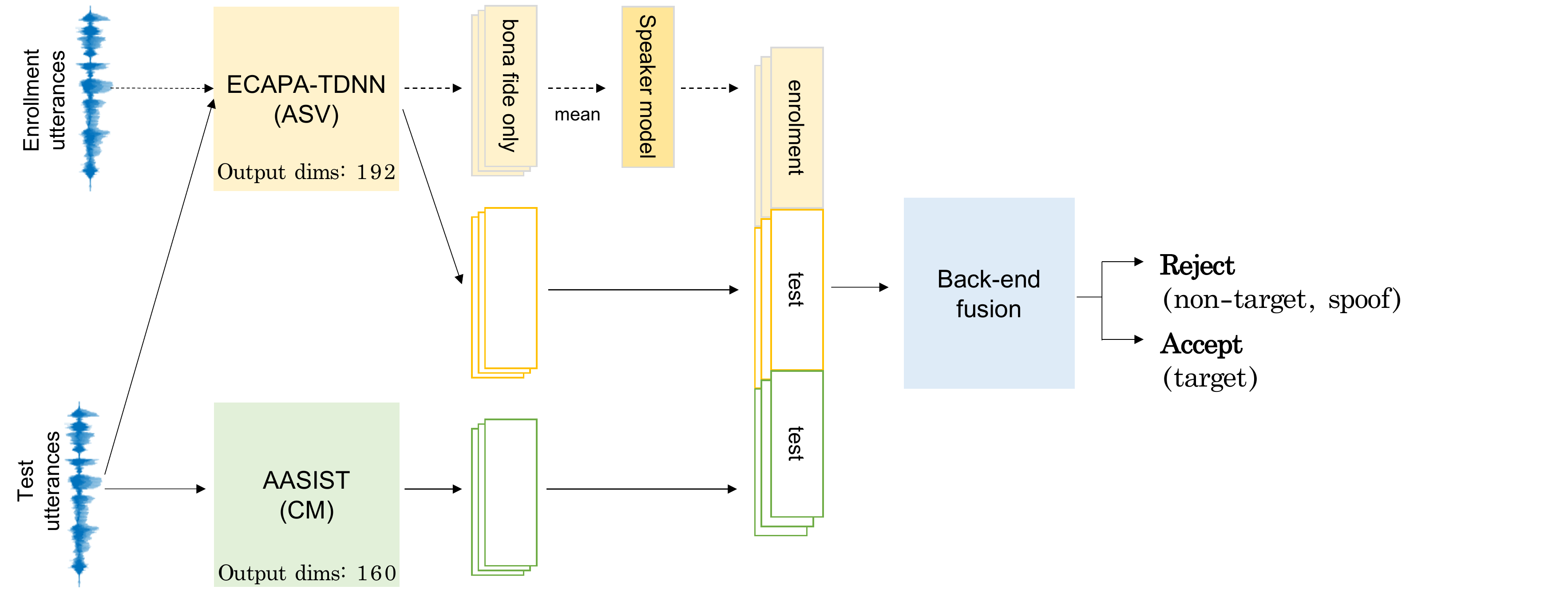}
 \caption{Illustration of the back-end DNN fusion. Three embeddings are fed to the DNN; only a speaker embedding is extracted from enrolment utterance and both speaker and spoofing embeddings are extracted from test utterance. In the training phase, `mean' is removed because only one enrolment utterance is involved. In both the development and evaluation phases, there exist multiple enrolment utterances, hence, embeddings are averaged element-wise before being fed to the DNN.}
   \label{fig:dnn_fusion}
 \end{figure*}

Our DNN-based baseline system, depicted in Fig.~\ref{fig:dnn_fusion} and introduced in Section~\ref{sec:ensemble}, is inspired by the solution in~\cite{shim2020integrated} and is based upon an ensemble of three embeddings.
The first and second are speaker (ASV) embeddings extracted from enrolment and test utterances respectively.  
The third is a spoofing (CM) embedding extracted only from the test utterance.
Described in this section are the backbone models used for their embedding extraction.

\subsection{Speaker embedding}
\label{ssec:spk_embd}
Driven by the availability of massive datasets, e.g.\ VoxCeleb~\cite{voxceleb2}, and competitive challenges, the performance of ASV systems has improved substantially in recent years~\cite{snyder2018x,desplanques2020ecapa}. 
The majority of today's best-performing ASV systems utilise some form of speaker embedding in a latent space in which linear classifiers can be applied (e.g., cosine similarity, probabilistic LDA).
We use the ECAPA-TDNN\footnote{\label{fn:ecapa-tdnn}\url{https://github.com/TaoRuijie/ECAPATDNN}} speaker embedding extractor, one of the most popular models in the recent ASV literature~\cite{desplanques2020ecapa}. 
It consists of a Res2net backbone architecture~\cite{gao2019res2net} with squeeze-excitation (SE) modules~\cite{hu2018squeeze}.
The model operates upon cepstral acoustic features and uses three SE-Res2net blocks where three-block outputs are all concatenated. 
Concatenated frame-level embeddings are then aggregated into a single utterance-level embedding leveraging an attentive statistical pooling (ASP) layer. 
The ASP layer is a variant of the original reported in~\cite{okabe2018attentive} which is also dependent on channel and context statistics among frame-level embeddings.
192-dimensional speaker embeddings are obtained by applying an affine transform with a fully-connected layer to the ASP layer output.
The model is trained using an additive angular margin softmax (AAM-softmax) objective function~\cite{deng2019arcface}.
Further details are available in~\cite{desplanques2020ecapa}.

\subsection{Spoofing embedding}
\label{ssec:spf_embd}

Progress in spoofing detection has been led by the ASVspoof initiative\footnote{\url{https://www.asvspoof.org/}} and associated challenge series~\cite{ASV2021challenge} which provides for benchmarking using common datasets, protocols and metrics.
The state-of-the-art methods apply end-to-end (E2E) DNNs with diverse architectures and strategies such as graph neural networks and graph attention networks (GATs)~\cite{tak2021end,jung2022aasist}. 

We use the E2E AASIST\footnote{\label{fn:assist}\url{https://github.com/clovaai/aasist}} spoofing detection model which delivers state-of-the-art performance for the ASVspoof 2019 LA database~\cite{jung2022aasist}. 
It operates directly upon raw waveform inputs using a variant of the RawNet2 encoder~\cite{jung2020improved} to generate three-dimensional feature maps (channel, spectral, and temporal). 
Two different views, (channel, spectral) and (channel, temporal), are then composed by applying an element-wise maximum operation to spectral and temporal axes.
The two views are further processed using graph modules that consist of GATs and a graph pooling layer.
Specially designed heterogeneous graph attention layers and max graph operations are then used to integrate the two views thereby combining temporal and spectral cues.  
A two-class prediction output is finally generated using a readout operation comprising a hidden fully connected (FC) layer.  
160 dimensional spoofing embeddings are extracted immediately prior to the FC output layer.
Further details are available in~\cite{jung2022aasist}.

\section{SASV}
\label{sec:ensemble}
Described in this section are three different approaches to combine CM and ASV sub-systems.  
They involve: i)~score-sum fusion; ii)~non-linear embedding fusion using DNNs; and iii)~cascaded combination.  
Each of the three strategies is described in the following.

\subsection{B1: Score-sum fusion}
\label{ssec:score_sum}
\vspace{-5pt}
The simplest strategy to system combination involves a straightforward score-sum. 
score-sum fusion is used widely and needs neither training nor fine-tuning. 
Our approach is based upon the ASV cosine similarity score (derived from enrolment and test utterances) and the CM output score (derived from the test utterance). 
ASV scores are calculated using cosine similarity with the range of -1 to 1.
CM scores are the softmax non-linearity outputs with the range of 0 to 1.
The score-sum back-end fusion serves as baseline1 (B1) for the SASV 2022 Challenge.\footnote{\label{fn:b1} Performance is different to that stated in the evaluation plan~\cite{jung2022sasv} on account of the softmax applied to the CM output which improves results notably.}


\subsection{B2: DNN back-end fusion}
The DNN-based fusion strategy is illustrated in Fig.~\ref{fig:dnn_fusion}.
It operates upon the set of three embeddings described in Section~\ref{sec:embd_extract}:   
a pair of speaker (ASV) embeddings extracted from enrolment and test utterances; 
a spoofing (CM) embedding extracted from the test utterance alone. 
As illustrated in Fig.~\ref{fig:dnn_fusion}, ASV embeddings and the CM embedding are combined using back-end DNN fusion.
The DNN back-end fusion serves as baseline2 (B2) for the SASV 2022 Challenge.
The DNN model contains three fully
connected layers with leaky ReLU non-linear activation functions. 
The last layer consists of two
neurons which correspond to two classes: (i)~target, (ii) non-target / spoof.
Element-wise average speaker embeddings are used in the case of multiple enrolment utterances.

\subsection{Cascaded combination}
\label{ssec:cascade}
Though not an SASV baseline, a cascaded combination is also reported here for reference and 
is the same approach as illustrated in Fig.~\ref{fig:three_cascade}-(a).
In practice, we utilise separate CM and ASV thresholds set using development data to make false acceptance rates and false rejection rates equal for both sub-systems.
Separate thresholds, optimised for the CM and ASV systems using the development protocol, are then applied without modification to evaluation data. 
The gate applied in the cascaded approach effectively combines CM and ASV systems at the decision level.  
While the CM produces a score, the ASV system produces scores only for trials labelled by the CM as bona fide (the gated decision).
Only decisions, not scores, are made consistently for each trial and,
as a result, there is no straightforward way to estimate the EER.
Thus, to estimate performance in the case of cascaded combination, we resort to error counting and estimation of the half-total error rate (HTER).
Given that we use the EER for the two baseline systems, but the HTER for the cascaded combination, we stress that results are not comparable.


\begin{table}[t]
  \caption{Description of EERs. The system involves enrolment utterance(s) and a test utterance. The enrolment utterance(s) is bona-fide (i.e. genuine) and test utterance belongs to either of the three types.}
  \centering
  \label{tab:eer_types}
  \begin{tabular}{lccc}
    \hline
    & Target & Non-target & Spoof\\
    \hline
    SV-EER & + & - &  \\
    SPF-EER & + &  & - \\
    SASV-EER & + & - & - \\
    \hline
  \end{tabular}
\end{table}

\section{Databases and protocols}
\label{sec:database}
We describe the databases used in this work: (i)~the VoxCeleb2 database~\cite{voxceleb2} used for training the 
ASV system; (ii)~the ASVspoof 2019 LA database~\cite{wang2020asvspoof} used for training the 
CM system and for SASV assessment.

\subsection{VoxCeleb2}

The ECAPA-TDNN model used to extract speaker embeddings was trained using the development partition of the VoxCeleb2 database. 
The dataset was collected by crawling online videos of celebrity interviews. 
Its development partition includes data collected from 5,994 speakers of which 61\% are male and 39\% are female. 
Network inputs are 80-dimensional mel filterbank acoustic features.
The network was trained following the recipe described in~\cite{ecapatdnn_pretrained}, where data augmentation is based on use of the room impulse response (RIR) database~\cite{rir} and additive noise recordings contained in the MUSAN database~\cite{musan}. 

\begin{table*}[ht]
\caption{
  The three different EERs (\%) for the SASV 2022 development and evaluation partitions.
  SASV-EER for all baselines are calculated using the entire protocol that includes trials used to measure the SV-EER (target vs.\ non-target) and those used to measure the SPF-EER (target vs.\ spoof).
  Results are shown for a conventional ASV system (ECAPA-TDNN) and the two baseline solutions. 
  B1 and B2 are baseline systems used for the SASV challenge}
\vspace{2mm}
\centerline{
\renewcommand{\arraystretch}{1.4}
\begin{tabular}{lcccccc}
\hline

\multirow{2}{*}{} & \multicolumn{2}{c}{SV-EER}  & \multicolumn{2}{c}{SPF-EER} & \multicolumn{2}{c}{SASV-EER}\\ 
\cline{2-7}
                  & \multicolumn{1}{c}{Dev} & \multicolumn{1}{c}{Eval} &  \multicolumn{1}{c}{Dev} & \multicolumn{1}{c}{Eval}& \multicolumn{1}{c}{Dev} & \multicolumn{1}{c}{Eval}               \\ 
\hline

ECAPA-TDNN & 1.88&1.63 & 20.30&30.75&17.38&23.83\\


\textbf{B1\textsuperscript{\ref{fn:b1}}}: Score-sum &1.99 &1.66&0.23&1.76&1.01&1.71\\

\textbf{B2}: DNN fusion &12.87&11.48&0.13&0.78&4.85&6.37\\
\hline
\end{tabular}}
\label{tab:baseline}
\end{table*}

\begin{table}[ht]
\caption{
  HTERs (\%) of the cascade solution for the SASV 2022 development and evaluation partitions. Thresholds that equals false acceptance and false rejection rates for each ASV and CM system is adopted (i.e., the ones that are used to measure EER).}
\vspace{2mm}
\centerline{
\begin{tabular}{lcccccc}
\hline

\multirow{2}{*}{} & \multicolumn{2}{c}{SV-HTER}  & \multicolumn{2}{c}{SPF-HTER} & \multicolumn{2}{c}{SASV-HTER}\\ 
\cline{2-7}
                  & \multicolumn{1}{c}{Dev} & \multicolumn{1}{c}{Eval} &  \multicolumn{1}{c}{Dev} & \multicolumn{1}{c}{Eval}& \multicolumn{1}{c}{Dev} & \multicolumn{1}{c}{Eval}               \\ 
\hline

Cascade &1.90&1.60&0.99&1.40&1.18&1.47\\

\hline
\end{tabular}}
\label{tab:Cascade results}
\end{table}

\subsection{ASVspoof 2019}
CM experiments were performed following the standard ASVspoof 2019 LA CM protocol described in~\cite{wang2020asvspoof}.  
It consists of disjoint train, development, and evaluation partitions. 
Each partition contains both bona fide and spoofed utterances where the latter are generated using 19 VC and TTS algorithms (6 for the train and development sets, 13 for the evaluation set). 
SASV evaluation is performed using the ASVspoof 2019 LA ASV protocol. 
The ASV protocol is not used by ASVspoof participants and is, instead, used only by the ASVspoof organisers to estimate ASV performance and tandem CM+ASV performance using the min t-DCF metric.  
For the SASV challenge, the ASV protocol is used by participants for experimentation involving
three different trials:  
\begin{enumerate}
    \item {\bfseries target} bona fide trials uttered by the same speaker as the  enrolment utterance(s); \vspace{-5pt}
    \item {\bfseries non-target} (zero-effort impostor) bona fide trials uttered by a different speaker as the enrolment utterance(s); \vspace{-5pt}
    \item {\bfseries spoofed} trials which are synthesised or converted to spoof the voice of the speaker in the enrolment utterance(s).
\end{enumerate}

Both development and evaluation protocols are provided with the freely available ASVspoof 2019 LA dataset\footnote{\url{https://datashare.ed.ac.uk/handle/10283/3336}} or with the open-source SASV baseline implementations.\footnote{\label{fn:baseline}\url{https://github.com/sasv-challenge/SASVC2022_Baseline}}
\begin{itemize}
    \item development protocol: \texttt{ASVspoof2019.LA.\\asv.dev.gi.trl.txt}; \vspace{-5pt}
    \item evaluation protocol: \texttt{ASVspoof2019.LA.\\asv.eval.gi.trl.txt}.
\end{itemize}


\section{Metrics}
\label{ssec:metrics}
We use the classical EER (SASV-EER) as the primary metric.
In keeping with the metrics used in~\cite{sahidullah2016integrated,todisco2018integrated}, the SASV-EER does not distinguish between non-target/zero-effort impostor and spoofed access attempts. 
Additional insights into performance can be gained from comparisons between the SASV-EER and: (i)~more traditional estimates of speaker verification performance (SV-EER) estimated from a set of target and non-target bona fide trials; (ii)~estimates of performance when non-target trials are replaced with spoofed trials (SPF-EER). 

Table~\ref{tab:eer_types} illustrates the trials types and ground-truth labels used to measure each of the three different EERs. 
As shown, all three EERs are estimates of ASV performance, with both the SV-EER and SPF-EER being estimated using different subsets of the full set of trials that are used for estimating the SASV-EER. 

\begin{table*}[!t]
    \centering
    \small
    \caption{
      Breakdown of SPF-EER (\%) and their pooled (P) EER for all 13 different spoofing attacks in the ASVspoof 2019 LA evaluation set, measured using SASV protocol without non-target trials.
      }
    \setlength\tabcolsep{4.7pt}
    \renewcommand{\arraystretch}{1.3}
    \begin{tabularx}{\linewidth}{l || *{13}{c}|| Y}
      \hline
      \textbf{System} &\textbf{A07}&\textbf{A08}&\textbf{A09}&\textbf{A10}&\textbf{A11}&\textbf{A12}&\textbf{A13}&\textbf{A14}&\textbf{A15}&\textbf{A16}&\textbf{A17}&\textbf{A18}&\textbf{A19}&  \textbf{P} 	\\ 
       \hline
       ECAPA-TDNN & 32.66 & 18.80 & 2.20 & 50.61 & 47.08 & 39.56 & 11.62 & 35.39 & 36.54 & 60.71 & 1.85 & 2.38 & 4.77 & 30.75 \\
     \textbf{B1\textsuperscript{\ref{fn:b1}}}:Score-sum & 2.05  & 0.69  & 0.07 & 7.19  & 0.32  & 4.42  & 0.07  & 0.08  & 1.75  & 1.18  & 0.73 & 1.18 & 0.63 & 1.76 \\
     \textbf{B2}:DNN fusion & 0.50 & 0.34 & 0.00 & 1.28 & 0.20 & 1.20 & 0.16 & 0.10 & 0.55 & 0.85 & 0.77 & 1.87 & 0.42 & 0.78\\    
    \hline
       
         
\hline
    \end{tabularx}
    \label{tab:breakdown}
\end{table*}
\section{Experiments}
\label{sec:exp}

We describe specific implementation details relating to embedding extraction, DNN-based fusion and hardware, followed by the presentation of experimental results.

\subsection{Implementation details}
\label{ssec:implementation_details}

\textbf{Speaker and spoofing embeddings} --
We used open source implementations for both the ECAPA-TDNN\textsuperscript{\ref{fn:ecapa-tdnn}} and AASIST\textsuperscript{\ref{fn:assist}}
models described in Sections~\ref{ssec:spk_embd} and~\ref{ssec:spf_embd}, respectively.
We used pre-trained weight parameters making embedding extraction fully reproducible.

\newpara{DNN-based back-end fusion, B2:}
We used a simple multi-layer perceptron for the back-end fusion with three hidden layers comprising 256, 128 and 64 nodes respectively, without regularisation (e.g., dropout, batch normalisation, and weight decay).
The DNN model for back-end fusion is trained on the ASVspoof 2019 LA train partition. The Adam optimiser was applied, and the learning rate was scheduled with warm starts between 0.1 to 0.001~\cite{loshchilov2017sgdr}.
The output node indicates whether the utterance is a target or not (non-target or spoof).
The fusion model is also publicly available.\textsuperscript{\ref{fn:baseline}}
%

\newpara{Hardware specification} --
All experiments reported in this paper were preformed using a single Nvidia 3090 GPU. The provided scripts can also be run on GPUs with less memory, e.g.\ an Nvidia 1080ti GPU.

\subsection{Results} 
\label{ssec:results}

Results are presented in Table~\ref{tab:baseline} which shows all three EERs for both development and evaluation partitions for the ECAPA-TDNN system and the two SASV baselines.  Results for the cascaded ensemble system in terms of HTERs are shown in Table~\ref{tab:Cascade results}.

\newpara{ECAPA-TDNN} -- Results for the ECAPA-TDNN are illustrated in the first row of Table~\ref{tab:baseline}. 
In view of the domain mismatch between ASVspoof data and the VoxCeleb2 data used for ASV training, the system performs reliably, with SV-EERs of 1.88\% and 1.63\% for the development and evaluation protocols respectively.
The SPF-EERs of 20.30\% and 30.75\% indicate that the same system is vulnerable to spoofing attacks.
The SASV-EERs, at 17.38\% and 23.83\%, are also high, confirming that the standalone ASV system provides little robustness to spoofing attacks.

\newpara{Score-sum fusion, B1} -- The second row of Table~\ref{tab:baseline} shows results using score-sum fusion described in Section~\ref{ssec:score_sum}.
SV-EER results are inline with results for the ECAPA-TDNN system. SPF-EER results, however, are substantially reduced.  Together, these results indicate the potential of the B1 baseline to improve robustness to both non-target trials and spoofing attacks, without impacts upon target trials. This finding is confirmed by SASV-EERs of 1.01\% and 1.71\% for the development and evaluation protocols.
The SASV-EER of 1.71\% remains 5\% higher relative to the ECAPA-TDNN SV-EER of 1.63\% implying that, while results are encouraging, impacts of spoofing remain.



\newpara{DNN fusion, B2} -- The last row of Table~\ref{tab:baseline} shows results for DNN-based back-end embedding-level fusion. 
With speaker and spoofing embeddings projected to a new representation space via an affine transformation, we expected DNN-based embedding-level fusion to outperform score-level fusion.
Relative to B1, the SPF-EER of 1.76\% is reduced by more than 50\% for the evaluation protocol.
However, the SV-EER increases from 1.66\% for B1 to 11.48\%, meaning that, with a vanilla multi-layer perception, the discriminative power of the speaker embedding is degraded in the joint representation space.
This explains the increase in the SASV-EER, to 6.37\%. 
%



\newpara{Cascade ensemble} -- Table~\ref{tab:Cascade results} shows results in terms of the HTER for cascaded CM and ASV systems as described in Section~\ref{ssec:cascade}. 
Lower error rates below the SV-HTER are observed for both the SPF-HTER and the SASV-HTER which drops 
to 1.47\%.
While error rates for the cascaded ensemble are lower than those for the two SASV baseline systems, results for the cascaded system are HTER estimates, not EER estimates.
Therefore, they are not comparable and should not be interpreted as meaning that the cascaded ensemble is the best approach.  
They might be interpreted, instead, as scope to improve SASV-EER results for the B1 and B2 baselines.

\newpara{Breakdown of SPF-EER} -- 
Table~\ref{tab:breakdown} shows a breakdown of SPF-EER results for each of the 13 different spoofing attacks, with pooled results to the right.
B2 outperforms B1 for the majority of attacks.  B1 outperforms B2 for A13, A14, A17, and A18 attacks, but the difference is almost negligible.
For B2, 
the SPF-EER is under 1\%
for all attacks except three. 
For A10 and A12, B1 SPF-EERs are 461\% and 268\% higher relative to B2 results. 
While B1 gives a lower SV-EER, B2 better harnesses the protection of the CM in deflecting spoofing attacks which hence leads to a lower SPF-EER.  It is the relative weakness in terms of the SV-EER that results in B2 having a higher SASV-EER.



%
%

\newpara{Further analysis.}
The trends shown in Table~\ref{tab:baseline} are unexpected; we expected better results for the DNN-based back-end. 
Using current ASV and CM sub-systems, simple fusions that operate at embedding or score levels give the best performance. 
However, we remain convinced in the merit of DNN-based embedding-level fusion approaches and predict that further research will reduce error rates considerably. 
B2 has better potential to harness the \emph{synergy} between CM and ASV sub-systems;
this is not the case for B1.
With advanced architectures, regularisation, and training strategies that better exploit the synergy, ASV and CM embedding fusion should give better performance and might even deliver SASV-EERs below the SV-EER.
We expect to observe such improvements in the forthcoming SASV 2022 Challenge. 
We argue that, with better potential for joint optimisation and hence better performance, future work should focus on the development of \emph{single, integrated} SASV solutions. 
Their development is the ultimate goal of the SASV challenge.

\section{Conclusions}
\label{sec:conclusion}
We found that, despite rapid advances in ASV, the state-of-the-art ECAPA-TDNN system remains vulnerable to spoofing attacks, hence, motivating either (i)~its combination with a standalone CM system or (ii)~the development of integrated spoofing aware ASV (SASV) solutions.
We explored different back-end integration (fusion) techniques, a straightforward score-sum fusion and a more sophisticated DNN-based approach. 
Surprisingly, the simple score-sum ensemble outperforms the DNN-based approach. 
This result may imply that simple back-end fusions which operate upon ASV and CM scores may be a sufficient solution.
Nonetheless, 
DNN-based back-end solutions and single integrated approaches
have greater potential for joint-optimisation to better exploit the synergy between CM and ASV solutions.
We hope to encourage this work through the SASV 2022 Challenge.  
In the end, there is only a \emph{single}, common task - reliable ASV -- a task that might best be solved with a \emph{single} SASV system.


\section{Acknowledgements}

The contributions of EURECOM authors is supported by the ExTENSoR project funded by the French Agence Nationale de la Recherche (ANR).

\clearpage
\bibliographystyle{IEEEbib}
\bibliography{Odyssey2022_BibEntries}
\end{document}